\documentclass[12pt]{iopart}
\usepackage{iopams,graphics,epsf}

\newcommand{\bmsigma}{\boldsymbol \sigma}

\newcommand{\bmA}{\boldsymbol A} 
\newcommand{\bmC}{\boldsymbol C}

\begin{document}

\title[Entanglement dynamics...]{Entanglement dynamics for two harmonic oscillators coupled to independent environments}

\author{Ruggero Vasile}

\address{Department of Physics and Astronomy University of Turku, FI-20014 Turun Yliopisto, Finland}
\ead{ruggero.vasile@utu.fi}
\begin{abstract}
We study the entanglement evolution between two harmonic oscillators
having different free frequencies each leaking into an independent
bath. We use an exact solution valid in the weak coupling limit and
in the short time non-Markovian regime. The reservoirs are identical
and characterized by an Ohmic spectral distribution with
Lorents-Drude cut-off. This work is an extension of the case
reported in \cite{Vasile} where the oscillators have the same free
frequency.
\end{abstract}


\maketitle

\section{Introduction}

Entanglement is considered a fundamental feature and a resource in
the context of quantum computation and information physics
\cite{NieChu}. However entanglement between quantum systems is
easily destroyed when their unavoidable interaction with the
external environment is taken into account \cite{weiss}. Features
and timing of the disentanglement phenomenon depend strongly on the
particular physical system and environment under investigation.
However, even if disentanglement cannot be avoided, quantum
information and computation tasks can still be performed when the
disentanglement time is much longer than the time needed to run the
quantum task. From this point of view it becomes of fundamental
importance to investigate mechanisms and features of disentanglement
in those systems used in the context of quantum computation and
information.

Continuous variable (CV) quantum systems are some of the possible
candidates for quantum protocols \cite{Braunstein}. We consider here
the problem of two independent harmonic oscillators each one leaking
in a bosonic structured reservoir
\cite{PraBec2004,Xiang2008,SerIllPar2004,DoddHall2004,Dodd2004,Hiro2001,MaOliPa}.
This situation has been considered also by the present author in
\cite{Vasile}. However here, as an extension, we study the case of
oscillators with different frequencies, initially in an entangled
TWB state. Our aim is to investigate the entanglement evolution in
the case of two identical high-T Ohmic reservoir with a
Lorentz-Drude cutoff. In this kind of systems, whose dynamics is
determined by a generalized Hu-Paz-Zhang master equation
\cite{HuPazZhang}, all the initial Gaussian states remain Gaussian
during the evolution. For this reason we are entitled to use the
separability criteria for CV systems introduced by Simon
\cite{Simon}, as a marker of entanglement between the oscillators.

The paper is organized as follows. In Section II we introduce the
physical model and its master equation in the weak coupling limit.
In Section III we review the concept of two-mode Gaussian squeezed
state, or Twin Beam state, and provide the solution to the master
equation. In Section IV we introduce the separability criteria for
bipartite continuous variable systems and the Lorentz-Drude spectral
function. Moreover we study the entanglement dynamics as a function
of the initial amount of entanglement and the relative values of the
free oscillator frequencies and the spectral function cut-off
frequency. Finally in Sec. V we give a brief summary of our work.

\section{The master equation}
Our system consists of two non-interacting quantum harmonic
oscillators with frequencies $\omega_1$ and $\omega_2$. Each
oscillator is coupled to its own bosonic reservoir. The total
Hamiltonian can be written as
\begin{eqnarray}\label{Ham}
H&=&\sum_{j}\hbar\omega_j a^{\dag}_j a_j+\sum_{j,k}\hbar\omega_{jk}
b^{\dag}_{jk} b_{jk}+\sum_{j,k}
\gamma_{jk}(a_j+a_j^{\dag})(b_{jk}+b_{jk}^{\dag}),
\end{eqnarray}
where $j=1,2$,  $\omega_{1k}$ and $\omega_{2k}$ are the frequencies
of the reservoirs modes,  $a_{j}$ ($a_{j}^{\dag}$) and $b_{jk}$
($b_{jk}^{\dag}$) are the annihilation (creation) operators of the
system and reservoirs harmonic oscillators, respectively, and
$\gamma_{jk}$ describe the coupling between the $j$-th oscillator
and the $k$-th mode of its environment. We assume reservoirs with
the same spectral structure and equally coupled to the oscillators.
The dynamics of the two oscillators can be described through the
following non-Markovian local in time master equation
\cite{HuPazZhang}
\begin{eqnarray}\label{HuPaZang}
\dot{\rho}(t)=\sum_j\frac{1}{i\hbar}[H_j^{0},\rho(t)]-\Delta_j(t)
[X_j,[X_j,\rho(t)]]+\Pi_j(t)[X_j,[P_j,\rho(t)]]+\nonumber\\
\frac{i}{2}r_j(t)[X_j^2,\rho(t)]-i\gamma_j(t)[X_j,\{P_j,\rho(t)\}],
\end{eqnarray}
where $\rho(t)$ is the reduced density matrix, $H_j^{0}$ is the free
Hamiltonian of the $j$-th oscillator, and
$X_j=\frac{1}{\sqrt{2}}(a_j+a_j^{\dag}),\quad
P_j=\frac{i}{\sqrt{2}}(a_j^{\dag}-a_j)$ are the quadrature
operators. The interaction with the reservoirs is taken into account
through the time-dependent coefficients of Eq. \eref{HuPaZang}. The
quantities $\Delta_j(t)$ and $\Pi_j(t)$ describe diffusion
processes, $\gamma_j(t)$ is a damping term and $r_j(t)$ renormalizes
the free oscillator frequencies $\omega_j$. For environments in
thermal equilibrium and in the weak coupling limit in ($r_j(t)$
being negligible), they read
\begin{eqnarray}\label{Coeff1}
\Delta_j(t)&=\alpha^2\!\!\int_0^t\!\! ds \!
\int_0^{+\infty}\!\!\!\!\!\!\!\! \!d\omega
J(\omega)[2N(\omega)+1]\cos(\omega s)\cos(\omega_j s), \nonumber\\
\Pi_j(t)&=\alpha^2\!\!\int_0^t\!\! ds \!
\int_0^{+\infty}\!\!\!\!\!\!\!\! \!d\omega
J(\omega)[2N(\omega)+1]\cos(\omega s)\sin(\omega_j s), \nonumber\\
\gamma_j(t)&=\alpha^2\!\!\int_0^t\!\! ds \!
\int_0^{+\infty}\!\!\!\!\!\!\!\! \!d\omega J(\omega)\sin(\omega
s)\sin(\omega_j s).
\end{eqnarray}
\section{Gaussian states and separability condition}
As initial states for our system we consider the class of two-mode
squeezed states (or twin-beams TWB) obtained applying a two-mode
squeezing operator to the vacuum state of the oscillators
\cite{Braunstein}. They belong to the class of Gaussian states
characterized by a Gaussian characteristic function
$\chi_0(\Lambda)=\exp\bigl\{-\frac{1}{2}\Lambda^{T}\bmsigma_0\Lambda\bigl\}$,
where $\bmsigma_0$ is the covariance matrix and
$\Lambda^T=(x_1,p_1,x_2,p_2)$. For a TWB state we have
\begin{equation}
\bmsigma_0 = \left(
\begin{array}{c  c}
\bmA_{0} & \bmC_0 \\  \bmC_0^{T} & \bmA_0
\end{array}
\right),
\end{equation}
with $\mathbf{A}_0=Diag(a,a)$, $\mathbf{C}_0=Diag(c,-c)$,
$a=\cosh(2r)/2$ and $c=\sinh(2r)/2$. The TWB state is thus
determined only by the \emph{squeezing} parameter $r$ which also
determines the initial amount of entanglement. The bigger is $r$,
the larger is the initial entanglement. Using the characteristic
function solution \cite{Intra2003}, it is trivial to verify that the
characteristic function at time $t$ maintains its Gaussian character
$\chi_t(\Lambda)=\exp\bigl\{-\frac{1}{2}\Lambda^{T}\bmsigma_t\Lambda\bigl\}$,
where
\begin{equation}\label{CovMatT2}
\bmsigma_{t} = \left(
\begin{array}{c  c}
\bmA_{t}^{(1)} & \bmC_{t} \\  \bmC_{t}^{T} & \bmA_{t}^{(2)}
\end{array}
\right),
\end{equation}
with
\begin{equation}
\mathbf{A}_t^{(i)}=\mathbf{A_0}e^{-\Gamma_i}+\left(
                                         \begin{array}{cc}
                                           \Delta_{\Gamma}^{(i)}+\Delta_{co}^{(i)}-\Pi_{si}^{(i)} & -\Delta_{si}^{(i)}+\Pi_{co}^{(i)} \\
                                           -\Delta_{si}^{(i)}+\Pi_{co}^{(i)} & \Delta_{\Gamma}^{(i)}+\Delta_{co}^{(i)}-\Pi_{si}^{(i)} \\
                                         \end{array}
                                       \right),
\end{equation}
\begin{equation}
\mathbf{C}_t=\left(
                                         \begin{array}{cc}
                                           ce^{-(\Gamma_1+\Gamma_2)}\cos[(\omega_1+\omega_2)t] & ce^{-(\Gamma_1+\Gamma_2)}\sin[(\omega_1+\omega_2)t] \\
                                           -ce^{-(\Gamma_1+\Gamma_2)}\sin[(\omega_1+\omega_2)t] & ce^{-(\Gamma_1+\Gamma_2)}\cos[(\omega_1+\omega_2)t] \\
                                         \end{array}
                                       \right),
\end{equation}
with $\Gamma_i(t)=2\int_0^t\gamma(s)ds$. Moreover, because we are
interested in the short time non-Markovian dynamics only, we defined
$\Delta_{\Gamma}^{(i)}(t)\simeq\int_0^t\Delta_i(s)ds$, and the
following \emph{secular coefficients}
$\Delta_{co}^{(i)}(t)\simeq\int_0^t\Delta_i(s)\cos[2\omega_i(t-s)]ds$,
$\Delta_{si}^{(i)}(t)\simeq\int_0^t\Delta_i(s)\sin[2\omega_i(t-s)]ds$,
$\Pi_{co}^{(i)}(t)\simeq\int_0^t\Pi_i(s)\cos[2\omega_i(t-s)]ds$ and
$\Pi_{co}^{(i)}(t)\simeq\int_0^t\Pi_i(s)\sin[2\omega_i(t-s)]ds$.
Details of the calculations can be found in
\cite{Vasile,MaOliPa,Intra2003}. If a suitable environment spectrum
is provided, all previous coefficients can be evaluated analytically
in the high temperature limit. This is the case of the Lorentz-Drude
Ohmic distribution we introduce in the next section. In equations
(6) and (7) we omitted the explicit time dependence for all the
appearing coefficients.

\section{Entanglement dynamics}
In this section we investigate the entanglement dynamics between the
oscillators using the separability criterion of Simon \cite{Simon}.
This criterion is well-suited in the context of two-mode Gaussian
states because it represents a necessary and sufficient condition
for separability and it depends only on the analytic form of the
covariance matrix. In the case of the time-dependent and non
symmetric covariance matrix \eref{CovMatT2}, the Simon criteria is
equivalent to the following algebraical inequality \cite{PraBec2004}
\begin{eqnarray}
S(t)=\det[\mathbf{A}_t^{(1)}\mathbf{A}_t^{(2)}]
+(\frac{1}{4}-|\det\mathbf{C}_t|)^2-Tr[\mathbf{A}_t^{(1)}\mathbf{J}\mathbf{C}_t\mathbf{J}\mathbf{A}_t^{(2)}
\mathbf{J}\mathbf{C}_t^T\mathbf{J}]\nonumber\\
-\frac{\det[\mathbf{A}_t^{(1)}]+\det[\mathbf{A}_t^{(2)}]}{4}\geq 0
\end{eqnarray} with
\begin{equation}
\mathbf{J}=\left(
             \begin{array}{cc}
               0 & 1 \\
               -1 & 0 \\
             \end{array}
           \right).
\end{equation}
$S(t)$ is  called \emph{separability function}. When $S(t)< 0$ the
state is entangled, otherwise it is separable.

We now consider a particular situation with two reservoirs in
thermal equilibrium at high temperature $T$ ($\hbar\omega_c<<k_B T$)
characterized by a Lorentz-Drude \emph{Ohmic} spectral function
\cite{weiss}
\begin{equation}
J(\omega)=\frac{\omega_c^2}{\pi}\frac{\omega}{\omega^2+\omega_c^2}
\end{equation}
where $\omega_c$ is the cut-off frequency of the distribution.
Hitherto we fix the temperature through the condition
$k_BT/\hbar\omega_c=100$ and the coupling constant to the value
$\alpha=0.1$. Moreover we use a dimensionless time variable
$\tau=\omega_c t$ for the evolution of the separabilty. With these
choices we have only two free parameters $x_i=\omega_c/\omega_i$
($i=1,2$), namely \emph{resonance parameters}, describing the
relative positions of the free oscillator frequencies $\omega_i$
with respect to the spectral cut-off frequency $\omega_c$. In
\cite{Vasile} has been observed the existence of two different
resonance parameters regions characterized by different qualitative
and quantitative behavior of entanglement, $x_1=x_2\geq 1$ and
$x_1=x_2<<1$. Exploiting this result we investigate three different
regimes: 1) $x_1,x_2\geq 1$, 2) $x_1,x_2<<1$ and 3) $x_1<<1$,$
x_2\geq 1$.

A second degree of freedom for our analysis is the choice of the
amount of entanglement in the initial TWB state, given by the value
of squeezing parameter $r$. In principle we do not have any
limitation in the choice of the value of $r$, however in real
situations it is not possible to realize two-mode squeezed states
with $r\geq 2$ \cite{Braunstein}. On the other hand when
$r\simeq0.01$ the amount of entanglement is so small that, at least
in the high temperature limit, disentanglement is a very fast
process almost independent from $x_1$ and $x_2$. For these reasons
we restrict our investigation to the intermediate region of
$0.01\leq r\leq1$.

\begin{figure}[!]
\begin{center}
\includegraphics{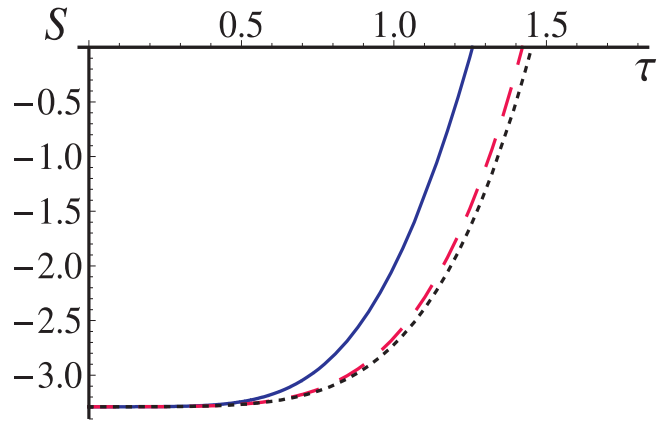}
\end{center}
\caption{(Colors online) Separability function $S$ vs scaled time
$\tau=\omega_c t$ with $r=1$ and $x_1=1$. The three curves are
relative to different values of $x_2$. The blue solid line
correspond to $x_2=1$, the red dashed line to $x_2=10$ and the black
dotted line to $x_2=100$.}\label{fig:1}
\end{figure}

We start the analysis considering the case $x_1,x_2\geq1$. We show
the entanglement evolution in Fig. \ref{fig:1} where we fixed the
squeezing parameter to the value $r=1$. In this regime the dynamics
is characterized by entanglement sudden death (ESD) appearing
already in the short time non-Markovian region for any value of $r$.
We also fixed the value of the parameter $x_1=1$ and plotted three
curves correspondent to $x_2=1$ (solid blue), $x_2=10$ (dashed red)
and $x_2=100$ (dotted black). We observe that the disentanglement
time $\tau_{dis}$, defined as $S(\tau_{dis})=0$, only slightly
increases as the adimensional frequency \emph{detuning} $\Delta
x=|x_1-x_2|$ increases.

\begin{figure}[!]
\begin{center}
\includegraphics{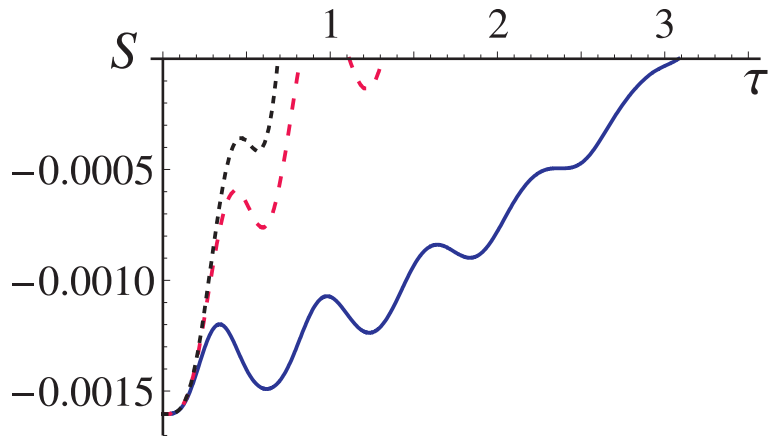}
\end{center}
\caption{(Colors online) Separability function $S$ vs scaled time
$\tau=\omega_c t$ with $r=0.04$ and $x_1=0.1$. The three curves are
relative to different values of $x_2$. The blue solid line
correspond to $x_2=0.1$, the red dashed line to $x_2=0.2$ and the
black dotted line to $x_2=0.3$.}\label{fig:2}
\end{figure}

The case $x_1,x_2<<1$ is reported in Fig. \ref{fig:2}. As emphasized
in detail in \cite{Vasile}, this regime is characterized by a more
variegate entanglement dynamics, showing oscillations, ESD and
revivals. We report the situation relative to an initially small
amount of entanglement correspondent to $r=0.04$ and $x_1=0.1$. The
three curves correspond to $x_2=0.1$ (solid blue), $x_2=0.2$ (dashed
red) and $x_2=0.3$ (dotted black). In all cases there are
entanglement oscillations and ESD, while an entanglement revival is
also present when $x_2=0.2$. As in the previous regime of Fig.
\ref{fig:1} we notice that disentanglement time is larger when
$\Delta x$ increases with fixed value of $x_1$. However here the
differences in the disentanglement time are much more evident. Also
oscillations are damped out as the detuning increases.

\begin{figure}[!]
\begin{center}
\includegraphics{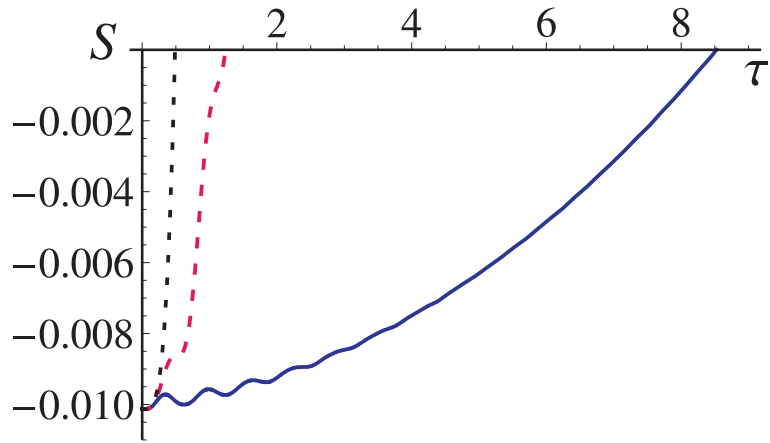}
\end{center}
\caption{(Colors online) Separability function $S$ vs scaled time
$\tau=\omega_c t$ with $r=0.1$. Blue solid line correspond to the
choice $x_1=x_2=0.1$, red dashed line to $x_1=0.1,x_2=1$, and black
dotted line to $x_1=x_2=1$}\label{fig:3}
\end{figure}

Finally we look at the regime $x_1<<1$ and $x_2\geq1$ whose results
are shown in Fig. \ref{fig:3} with the choice $r=0.1$. The blue
solid line corresponds to  $x_1=x_2=0.1$, the dotted black line
describes the $x_1=x_2=1$ case, and the red dashed line represents
the intermediate case of $x_1=0.1$ and $x_2=1$. Here we can observe
two completely different behaviors of $S(\tau)$ relative to the
cases $x_1=x_2=0.1$ and $x_1=x_2=1$. The former case is
characterized by evident entanglement oscillations and a much longer
survival time of entanglement. Instead the latter case shows a fast
ESD without signs of oscillations. The intermediate case of $x_1\neq
x_2$ possesses intermediate features: weak oscillations and a fast
ESD.

The three parameters regimes analyzed display all the possible
features in the entanglement dynamics for our particular system:
entanglement oscillations, ESD and revivals. Thus when the
oscillators frequencies are detuned no new qualitative behavior
emerges, compared to the equal frequency case \cite{Vasile}. On the
other hand we stress the existence of evident and non-negligible
differences. This does not happen in the regime $x_1,x_2>1$, where
the value of disentanglement time increases only slightly as the
oscillator detuning $\Delta x$ increases.

Quantitative differences characterize the opposite case, i.e., the
regime $x_1,x_2<<1$. Even a slight detuning ($\Delta x=0.1$) can
change drastically the value of the disentanglement time and damp
entanglement oscillations. Again this feature is present in Fig.
\ref{fig:3} where these discrepancies are even more pronounced.

From our investigation we can extract two main results. When the
oscillators have different frequencies ($\omega_1\neq\omega_2$), the
entanglement dynamics shows a behavior which is intermediate between
the cases of frequencies both equal to $\omega_1$ and both equal to
$\omega_2$. Moreover the leading role in the disentanglement process
seems to be played by the oscillator with higher resonance parameter
$x$, which forces the entanglement to evolve in a way similar to the
case of both parameters equal to the larger between the two, as
shown in Fig. \ref{fig:3}.

\section{Summary}
In this paper we extended the results obtained in \cite{Vasile} to
the case in which the two oscillators have different free
frequencies, thus when they interact with different parts of the
reservoirs spectra. We considered the particular case of an Ohmic
distribution in the high temperature limit for different initial TWB
states and different values of the free oscillator frequencies.
Quantitative differences in the entanglement dynamics can be
observed in particular when both resonance parameters are in the
regime $x<<1$, where the evolution changes also for a small values
of $\Delta x$. On the contrary the dynamics is not strongly affected
in the opposite regime even for large detuning. Moreover when the
two frequencies lay in opposite parameters regime, the dynamics is
similar to the case $x>>1$, characterized by lack of oscillations
and fast ESD.

\section*{Acknowledgements} The author wish to thank Dr S
Maniscalco, Dr S Olivares and Dr M G A Paris for fruitful
discussions about the present work. Magnus Ernhrooth foundation is
acknowledged for financial support.

\end{document}